\documentclass[aps,twocolumn]{revtex4}

\usepackage{booktabs}
\usepackage{setspace}
\usepackage{hyperref}
\usepackage{graphicx}% Include figure files
\usepackage{color}
\usepackage{epsfig}
\usepackage{subfig}
\usepackage{dcolumn}% Align table columns on decimal point
\usepackage{amsmath}
\usepackage{xcolor}
\usepackage{natbib}
\usepackage{caption}
\usepackage{bm}% bold math

\def\be{\begin{equation}}
\def\ee{\end{equation}}
\def\bea{\begin{eqnarray}}
\def\eea{\end{eqnarray}}
\def\NO{\nonumber}

\begin{document}

\title{Perturbative QCD Evidence for Spin-2 Particles in the Di-$J/\psi$ Resonances}% Force line breaks with \\

\author{Hong-Fei Zhang$^{a,b}$\footnote{Corresponding Author: shckm2686@163.com}, Yan-Qing Ma$^{c,d,e}$\footnote{Corresponding Author: yqma@pku.edu.cn}, Wen-Long Sang$^{f}$}
\affiliation{
a. College of Big Data Statistics, Guizhou University of Finance and Economics, Guiyang, 550025, China \\
b. Research Centre of Big-data Corpus \& Language Projects, School of Foreign Languages, Guizhou University of Finance and Economics, Guiyang, 550025, China \\
c. School of Physics and State Key Laboratory of Nuclear Physics and Technology, Peking University, Beijing 100871, China \\
d. Center for High Energy physics, Peking University, Beijing 100871, China \\
e. Collaborative Innovation Center of Quantum Matter, Beijing 100871, China \\
f. School of Physical Science and Technology, Southwest University, Chongqing 400700, China
}

\maketitle

Although tetraquarks and pentaquarks~\cite{GellMann:1964nj} were predicted along with the quark model in 1964,
more than half a century passed, multiquark-like states observed in experiment are still rare.
Among these states, the structure of hardly any was identified;
whether they are multiquarks, molecules of hadrons, or other possible species of resonances are still under debate
(for a review, see e.g. Refs.~\cite{Chen:2016qju, Liu:2019zoy}).

In 2020, a resonance of $J/\psi$ pair around 6.9 $\mathrm{GeV}$, $X$(6900), is observed by LHCb~\cite{LHCb:2020bwg},
and confirmed by the ATLAS~\cite{ATLAS:2023bft} and CMS~\cite{CMS:2023owd} Collaborations.
In addition, a new resonance at around $6.6~\mathrm{GeV}$ (we call it $X$(6600) in the rest of the paper) and an evidence at around $7.2~\mathrm{GeV}$ are also observed,
and a lower state around $6.2~\mathrm{GeV}$ ($X(6200)$), according to Ref.~\cite{Dong:2020nwy}, was indicated by the data.
All these states could be either tetraquarks ($cc\bar{c}\bar{c}$) or states consisting of two hidden-charm mesons bound with screened strong forces.
We call the latter a molecule-like state for convenience, regardless of the true binding mechanism of its constituent hadrons.
Here comes a crucial question:
What are the quantum numbers of the resonances?

The mass spectra of the fully-heavy exotic matter have been extensively studied using various models and techniques.
Unfortunately, there still does not exist any approach that can achieve high-precision prediction of hadron masses,
which limits the power of revealing the nature of the observed resonances.
For instance, the mass deviation of a spin-2 charmonium from its spin-0 companion is below $150~\mathrm{MeV}$, hence,
it is not reliable to identify the spin of a hidden-charm hadron by matching its mass to the theoretically obtained mass spectra
if the precision of the method is not reasonably below that value.

A good feature of the fully-heavy matter is that they consist only of heavy constituents,
hence their production cross sections can be handled within the framework of nonrelativistic QCD~\cite{Caswell:1985ui,Bodwin:1994jh},
which factorizes the cross section as a double expansion in both the strong coupling constant ($\alpha_s$) and the typical velocity ($v$) of the constituents inside the exotic states.
We will demonstrate that this framework provides a robust probe to the spin of the di-$J/\psi$ resonances.

The $J^{PC}$ of an $S$-wave hadron consisting only of charmed quarks and decaying exclusively into two $J/\psi$'s can only be $0^{++}$ or $2^{++}$.
The corresponding tetraquarks are denoted in the following as $T_{cc\bar{c}\bar{c}}(^1S_0)$ and $T_{cc\bar{c}\bar{c}}(^5S_2)$, respectively,
and molecule-like states as $M_{\psi\psi}(^1S_0)$ and $M_{\psi\psi}(^5S_2)$, respectively,
where the subscript $\psi$ stands for a $^3S_1$ charmonium.
Note here that the two components of a molecule-like state can be different from each other,
for example, they can be a $J/\psi$ and a $\psi'$,
and the corresponding molecule is $M_{J/\psi\psi'}$.

In nonrelativistic QCD~\cite{Caswell:1985ui,Bodwin:1994jh}, the production cross section of such an aforementioned exotic hadron $H$ can be factorized as
\bea
\mathrm{d}\sigma(H)=\sum_n\mathrm{d}\hat{\sigma}(n)\langle\mathcal{O}^H(n)\rangle, \label{eqn:cs}
\eea
where $\hat{\sigma}_n$ are perturbatively calculable short-distance coefficients (SDCs),
which are independent of the species of the produced hadron $H$,
$\langle\mathcal{O}^H\rangle$ are long-distance matrix elements (LDMEs),
which are independent of the short-distance process producing the charm quarks and antiquarks,
and the summation runs over all the allowed intermediate states with quantum numbers $n$.
Note here that the LDMEs for the molecule-like state production encapsulate those for the charmonia production.
Up to leading order (LO) in $v$, the intermediate states have the same components and quantum numbers as $H$,
and the corresponding LDME can be related to the normalized wave function of $H$ at the origin.
The higher-order corrections in $v$ comprise contributions from color-multiplet intermediate states and relativistic corrections.

This paper focuses exclusively on the color-singlet contribution for the following reason.
The color-multiplet contribution to the double $J/\psi$ hadroproduction becomes less significant in the region
where the invariant energy of the double $J/\psi$ system is close to twice the $J/\psi$ mass~\cite{He:2015qya},
specifically $m_{J/\psi J/\psi}\gtrsim2m_{J/\psi}$.
Meanwhile, the hadroproduction of the di-$J/\psi$ resonances involves the same topologies of Feynman diagrams as the double $J/\psi$ production,
leading to similar behavior in both cases.
This suggests that the color-multiplet contribution to the hadroproduction of the di-$J/\psi$ resonances is negligible compared to the color-singlet contribution.

In the color-singlet limit, Eq. (\ref{eqn:cs}) reduces to
\bea
\mathrm{d}\sigma(H)=\langle\mathcal{O}^{H}(n)\rangle\mathrm{d}\hat{\sigma}(n),
\eea
where $n$ has the same quantum numbers as $H$.
We present in Table~\ref{table:amc} the possible color and spin configurations of the intermediate states for each of the possible exotic mesons.
The intermediate states presented in the right-hand-side column are represented by their components, and color and spin configurations.
The first color and spin configurations in the parentheses are assigned to the systems consisting of the first and second components,
while the second to the third and fourth ones.
The colors and spins of the entire intermediate states are given in front of the parentheses.

\begin{table}
\caption{Possible color and spin configurations of the intermediate states for $S$-wave fully-charm meson production at LO in $v$.}
\begin{tabular}{c c}
\hline
$H$~&~$n$ \\
\hline
$T_{cc\bar{c}\bar{c}}(^1S_0)$~&~$cc\bar{c}\bar{c}[^1S_0^{[1]}(^1S_0^{[6]},^1S_0^{[\bar{6}]})]$,
$cc\bar{c}\bar{c}[^1S_0^{[1]}(^3S_1^{[\bar{3}]},^3S_1^{[3]})]$ \\
$T_{cc\bar{c}\bar{c}}(^5S_2)$~&~$cc\bar{c}\bar{c}[^5S_2^{[1]}(^3S_1^{[\bar{3}]},^3S_1^{[3]})]$ \\
$M_{\psi\psi}(^1S_0)$~&~$c\bar{c}c\bar{c}[^1S_0^{[1]}(^3S_1^{[1]},^3S_1^{[1]})]$ \\
$M_{\psi\psi}(^5S_2)$~&~$c\bar{c}c\bar{c}[^5S_2^{[1]}(^3S_1^{[1]},^3S_1^{[1]})]$ \\
\hline
\end{tabular}
\label{table:amc}
\end{table}

Note that two different intermediate states correspond to the spin-0 tetraquark.
There should also exist two different spin-0 $S$-wave tetraquark states.
We denote them as $T_{cc\bar{c}\bar{c}}(^1S_0)_s$ and $T_{cc\bar{c}\bar{c}}(^1S_0)_t$, respectively,
where the subscript $s$ ($t$) means the charm quark pair inside the tetraquark form a color-sextuplet (color-triplet) state.
The physical states should be two orthogonal linear combinations of $T_{cc\bar{c}\bar{c}}(^1S_0)_s$ and $T_{cc\bar{c}\bar{c}}(^1S_0)_t$.
We denote the one with larger hadroproduction cross section as $T_{cc\bar{c}\bar{c}}(^1S_0)$,
and it can be decomposed as $T_{cc\bar{c}\bar{c}}(^1S_0)=T_{cc\bar{c}\bar{c}}(^1S_0)_s\mathrm{sin}\theta+T_{cc\bar{c}\bar{c}}(^1S_0)_t\mathrm{cos}\theta$,
where $\theta$ is the corresponding mixing angle.
The other one is more difficult to observe thus is less likely to be one of the discovered resonances.

Note that the charm quark pair forming a color-sextuplet state repels each other,
which leads to the fact that the components inside $T_{cc\bar{c}\bar{c}}(^1S_0)_s$ are more loosely bound than those inside $T_{cc\bar{c}\bar{c}}(^1S_0)_t$.
Hence the wave function of $T_{cc\bar{c}\bar{c}}(^1S_0)_s$ at the origin should be much smaller than that of $T_{cc\bar{c}\bar{c}}(^1S_0)_t$.
Correspondingly, the color-sextuplet component of $T_{cc\bar{c}\bar{c}}(^1S_0)$ contribute little to the hadroproduction cross section, and we have
\bea
&&\mathrm{d}\sigma(T_{cc\bar{c}\bar{c}}(^1S_0))\approx\langle\mathcal{O}^{T_{cc\bar{c}\bar{c}}(^1S_0)_t}(cc\bar{c}\bar{c}[^1S_0^{[1]}(^3S_1^{[\bar{3}]},^3S_1^{[3]})])\rangle \NO \\
&&~~~~\times\mathrm{d}\hat{\sigma}(cc\bar{c}\bar{c}[^1S_0^{[1]}(^3S_1^{[\bar{3}]},^3S_1^{[3]})])\mathrm{cos}^2\theta.
\eea
By definition, the following inequality should be satisfied, $\mathrm{cos}^2\theta\ge0.5$.

For the above reasons, we neglect the contributions from the intermediate state $cc\bar{c}\bar{c}[^1S_0^{[1]}(^1S_0^{[6]},^1S_0^{[\bar{6}]})]$
and omit the color and spin configurations of the subsystems in the notation of the intermediate states.
For instance, $cc\bar{c}\bar{c}[^1S_0^{[1]}(^3S_1^{[\bar{3}]},^3S_1^{[3]})]$ will be designated as $cc\bar{c}\bar{c}[^1S_0]$.

The quantity that will be calculated in this paper is a ratio defined by
\bea
R(s)\equiv\frac{5\mathrm{d}\hat{\sigma}(s[^5S_2])}{\mathrm{d}\hat{\sigma}(s[^1S_0])},
\eea
where $s$ can be either $cc\bar{c}\bar{c}$ (for tetraquarks) or $c\bar{c}c\bar{c}$ (for molecules).
Evidently, it can be evaluated perturbatively.

The calculation is carried out up to LO in $\alpha_s$ and $v$.
Considering that the higher-order QCD corrections to quarkonia production are usually very important,
we need also to address this possibility in the di-$J/\psi$ resonance hadroproduction.
Its $p_t$ behavior at LO (in $\alpha_s$) is of next-to-leading power,
while at next-to-leading order, leading power (LP) contributions will emerge.
Besides this new behavior, the higher-order QCD corrections and the relativistic corrections are expected to be reasonable and proportional to the LO results,
thus are expected to be cancelled in the ratio.
In this work, we also calculate the LP contributions to the hadroproduction of the states studied in this paper,
and find that they start to overshoot the LO results at very high $p_t$,
at least higher than $30~\mathrm{GeV}$, which contribute very little to the integrated cross section concerned in the experiment.
We will also see later that the property of the LP part also support the main conclusion of this paper.
By tentatively varying the input parameters,
we find that the ratio is insensitive to the charm quark mass,
the parton distribution function choices, and the renormalization and factorization scales.

\begin{figure}
\begin{center}
\includegraphics*[scale=0.5]{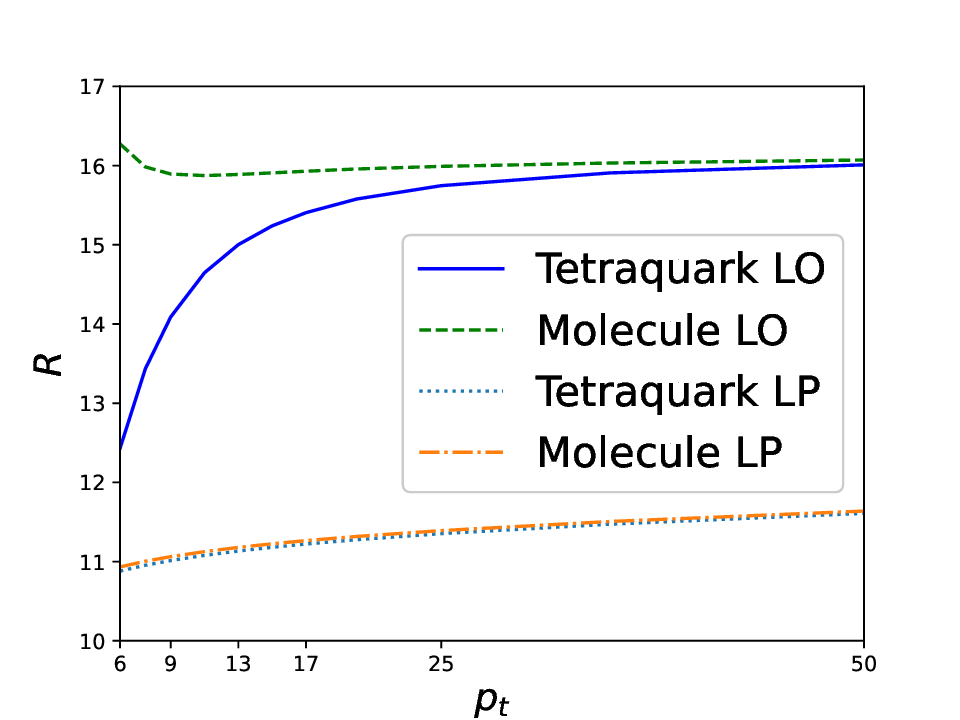}\\
\includegraphics*[scale=0.5]{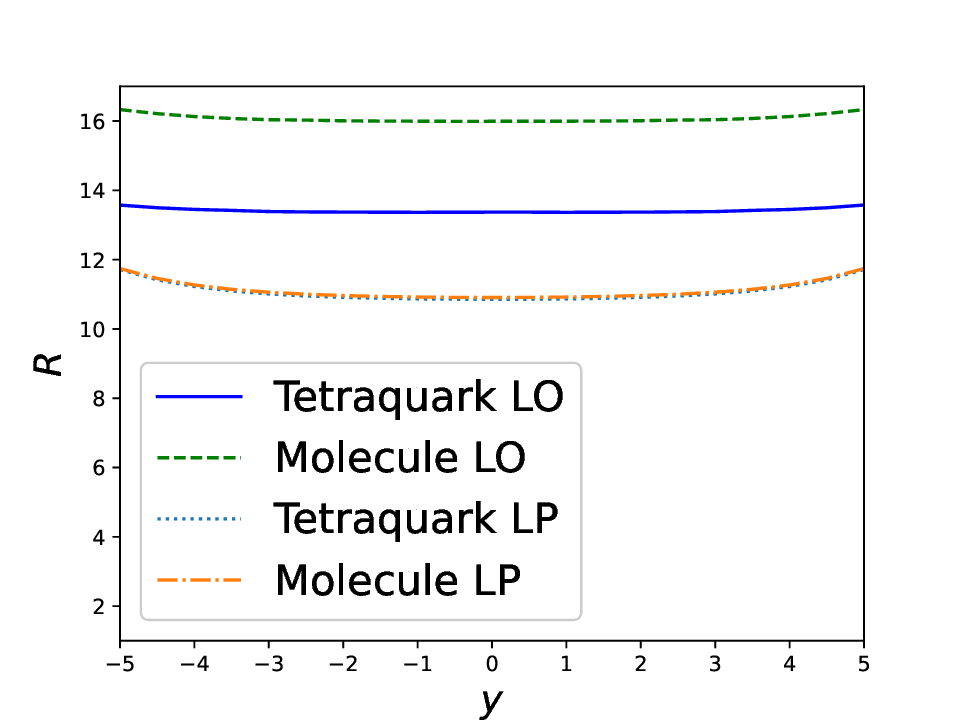}
\end{center}
\caption{
The ratio $R$ of the SDCs for spin-2 states to those for spin-0 ones as functions of $p_t$ and $y$.
}
\label{fig:Spin}
\end{figure}

The values of $R$ as functions of the transverse momentum ($p_t$) and rapidity ($y$) of the hadroproduced states are presented
independently for the LO perturbative results and LP contributions in Fig.~\ref{fig:Spin}.
We refer the readers to Refs.~\cite{Feng:2020riv, Feng:2023agq, Zhang:2023ffe} for the detail of the calculation framework.
The colliding energy is set to be $13~\mathrm{TeV}$ to accord with the experiments observing the di-$J/\psi$ resonances.
We can see that the ratio is almost independent of the kinematics of the intermediate states especially in the region $p_t\ge20~\mathrm{GeV}$.

By exploiting the heavy quark spin symmetry which leads to the following relations,
$\langle\mathcal{O}^{T_{cc\bar{c}\bar{c}}}(cc\bar{c}\bar{c}[^5S_2])\rangle\approx5\langle\mathcal{O}^{T_{cc\bar{c}\bar{c}}}(cc\bar{c}\bar{c}[^1S_0])\rangle$
and $\langle\mathcal{O}^{M_{\psi\psi}}(c\bar{c}c\bar{c}[^5S_2])\rangle\approx5\langle\mathcal{O}^{M_{\psi\psi}}(c\bar{c}c\bar{c}[^1S_0])\rangle$,
we have $\mathrm{d}\sigma(T_{cc\bar{c}\bar{c}}[^5S_2])/\mathrm{d}\sigma(T_{cc\bar{c}\bar{c}}[^1S_0])\approx R(cc\bar{c}\bar{c})/\mathrm{cos}^2\theta$,
and $\mathrm{d}\sigma(M_{\psi\psi}[^5S_2])/\mathrm{d}\sigma(M_{\psi\psi}[^1S_0])\approx R(c\bar{c}c\bar{c})$,
which implies that the perturbatively calculable ratio $R$ can well reflect the ratios of the cross sections.
Considering the range of $\mathrm{cos}^2\theta$,
we further have $R(cc\bar{c}\bar{c})\lesssim\mathrm{d}\sigma(T_{cc\bar{c}\bar{c}}[^5S_2])/\mathrm{d}\sigma(T_{cc\bar{c}\bar{c}}[^1S_0])\lesssim2R(cc\bar{c}\bar{c})$.
For both molecule-like states and tetraquarks,
the SDCs of the spin-2 states are much larger than those of their spin-0 counterparts.
The ratio is 16 for molecules and ranges from 12.5 to 16 for tetraquarks.
Considering the relations between $R$ and the ratio of the cross sections,
we assert that, in the region $6~\mathrm{GeV}\le p_t\le50~\mathrm{GeV}$ where the resonances are observed,
the hadroproduction cross sections of the spin-2 tetraquarks and molecules are much greater than their spin-0 counterparts.

Since the mass deviation of the two spin configurations is at the order of $m_cv^4$ ($v^2\approx0.3$ for fully-charmed hadrons),
approximately 100 $\mathrm{MeV}$,
it is expected that at least two resonances appear in a small energy interval around 6.9 $\mathrm{GeV}$ (as well as around 6.6 $\mathrm{GeV}$).
However, only one narrow resonance is observed at each specific mass.
A possible explanation is that the numbers of events of the other ones are too small and escaped the experimental resolution.
Thus, the observed one should have larger production cross section.
Note that 100 $\mathrm{MeV}$ is comparable with the width of the di-$J/\psi$ resonances,
which implies that other potential state(s) might hide inside the observed peaks or just inside the background.
As the data accumulates, it (they) might be recognized.
According to our calculation, the observed resonances are most likely spin-2 particles.

All our discussions are based on an assumption that the resonances are $S$-wave states.
We need also to address the possibility that it may be not.
In this case, the ground state might be below the di-$J/\psi$ threshold,
otherwise, it should also decay into a $J/\psi$ pair with a larger production cross section, which is detectable in experiment,
since the $P$-wave fully-heavy meson production cross section is suppressed by a factor of $v^2$ relative to its $S$-wave counterpart.

\begin{acknowledgments}
We thank Wei Chen for many useful communications and discussions.
This work is supported in part by the National Natural Science Foundation of China (12375079, 12325503, 11975029, 11965006, 11875071),
the National Key Research and Development Program of China (2020YFA0406400), 
the Guizhou Provincial Science and Technology Project (QKH-Basic-ZK[2021]YB319),
and the High-performance Computing Platform of Peking University.
Wen-Long Sang is supported by the Natural Science Foundation of ChongQing (CSTB2023 NSCQ-MSX0132).
\end{acknowledgments}

\vspace{0.3in}

\noindent\textbf{Author Contributions}

Hong-Fei Zhang and Yan-Qing Ma constructed the nonrelativistic QCD framework for the calculation of
the production of the fully-heavy tetraquarks and molecules consisting of two quarkonia used in this paper,
and independently performed the calculation of the leading-order results.
Wen-Long Sang calculated the results for the leading power contributions.

\end{document}